%% file: main.tex
\documentclass[twocolumn, superscriptaddress]{revtex4-1}

\usepackage{xspace}
\usepackage[colorlinks=true, allcolors=blue]{hyperref}
\usepackage[nolist]{acronym}
\usepackage{graphicx}
\usepackage{tabularx}
\usepackage{subfigure} 
\usepackage{siunitx}
\usepackage{multirow}
\usepackage{adjustbox}
\usepackage[normalem]{ulem}
\usepackage{booktabs}
\usepackage[table]{xcolor}
\DeclareSIUnit\msun{M_{\odot}}
\usepackage{mathtools}
\providecommand{\acrolowercase}[1]{\lowercase{#1}}

\begin{document}

\title{A neural network for estimating compact binary coalescence parameters of gravitational-wave events in real time}
\author{Sushant Sharma Chaudhary}
\email{sscwrk@mst.edu}
\affiliation{Institute of Multi-messenger Astrophysics and Cosmology, Missouri University of Science and Technology, Physics Building, 1315 N. Pine St., Rolla, MO 65409, USA}
\author{Gianmarco Puleo}
\affiliation{Physics Department, University of Trento, via Sommarive 14, I-38123 Trento, Italy}
\affiliation{INFN-TIFPA Trento Institute for Fundamental Physics and Applications, Via Sommarive, 14, 38123 Trento, Italy}
\affiliation{Scuola Internazionale Superiore di Studi Avanzati, Via Bonomea 265, 34136 Trieste, Italy}
\author{Marco Cavaglià}
\affiliation{Institute of Multi-messenger Astrophysics and Cosmology, Missouri University of Science and Technology, Physics Building, 1315 N. Pine St., Rolla, MO 65409, USA}

\begin{abstract}
Low-latency pipelines analyzing gravitational waves from compact binary coalescence events rely on matched filter techniques. Limitations in template banks and waveform modeling, as well as non-stationary detector noise cause errors in signal parameter recovery, especially for events with high chirp masses. We present a quantile regression neural network model that provides dynamic bounds on key parameters such as chirp mass, mass ratio, and total mass. We test the model on various synthetic datasets and real events from the LIGO-Virgo-KAGRA gravitational-wave transient GTWC-3 catalog. We find that the model accuracy is consistently over 90\% across all the datasets. We explore the possibility of employing the neural network bounds as priors in online parameter estimation. We find that they reduce by 9\% the number of likelihood evaluations. This approach may shorten parameter estimation run times without affecting sky localizations.
\end{abstract}
\maketitle

\begin{acronym}
\acro{TPR}[TPR]{True Positive Rate}
\acro{FPR}[FPR]{False Positive Rate}
\acro{BNS}[BNS]{binary neutron star}
\acro{NSBH}[NSBH]{neutron star-black hole}
\acro{MDC}[MDC]{Mock Data Challenge}
\acro{EoS}[EoS]{equation of state}
\acro{NS}[NS]{neutron star}
\acro{GW}[GW]{gravitational-wave}
\acro{GWs}[GWs]{gravitational waves}
\acro{LIGO}[LIGO]{Laser Interferometer Gravitational-Wave Observatory}
\acro{aLIGO}[aLIGO]{Advanced \acs{LIGO}}
\acro{AdVirgo}[AdVirgo]{Advanced Virgo}
\acro{O4}[O4]{fourth observing run}
\acro{LVK}[LVK]{LIGO, Virgo, and KAGRA}
\acro{LV}[LV]{LIGO-Virgo}
\acro{O1}[O1]{first observing run}
\acro{O2}[O2]{second observing run}
\acro{O3}[O3]{third observing run}
\acro{oLIB}[\acrolowercase{o}LIB]{Omicron+\acl{LIB}}
\acro{KAGRA}[KAGRA]{KAmioka GRAvitational\nobreakdashes-wave observatory}
\acro{GraceDB}[\texttt{GraceDB}]{GRAvitational-wave Candidate Event DataBase}
\acro{LLAI}[LLAI]{Low-Latency Alert Infrastructure}
\acro{CBC}[CBC]{compact binary coalescence}
\acro{BBH}[BBH]{binary black hole}
\acro{FAR}[FAR]{false alarm rate}
\acro{GstLAL}[GstLAL]{GStreamer LIGO Scientific Collaboration Algorithm Library}
\acro{SVD}[SVD]{singular value decomposition}
\acro{SNR}[SNR]{signal\nobreakdashes-to\nobreakdashes-noise ratio}
\acro{FGMC}[FGMC]{Farr, Gair, Mandel and Cutler formalism of Poisson Mixture Model}
%\acro{PyCBC}[PyCBC]{Python based Compact Binary Coalescence search software}
\acro{MBTA}[MBTA]{Multi-Band Template Analysis}
\acro{SPIIR}[SPIIR]{Summed Parallel Infinite Impulse Response}
\acro{GPU}[GPU]{Graphical Processing Units}
\acro{cWB}[\acrolowercase{c}WB]{Coherent WaveBurst}
\acro{GCN}[GCN]{General Coordinates Network}
\acro{SCiMMA}[SCiMMA ]{Scalable CyberInfrastructure for Multi-Messenger Astrophysics}
\acro{BAYESTAR}[\texttt{BAYESTAR}]{BAYESian TriAngulation and Rapid localization}
\acro{FITS}[FITS]{Flexible Image Transport System}
\acro{HEALPix}[HEALP\acrolowercase{ix}]{Hierarchical Equal Area isoLatitude Pixelization}
\acro{MOC}[MOC]{Multi-Order Coverage}
\acro{EM}[EM]{electromagnetic}
\acro{NS}[NS]{neutron star}
\acro{BH}[BH]{black hole}
\acro{LLOID}[LLOID]{Low Latency Online Inspiral Detection}
\acro{IIR}[IIR]{Infinite Impulse Response}
\acro{ToO}[ToO]{Target of Opportunity}
\acro{PE}[PE]{Parameter Estimation}
\acro{Bilby}[Bilby]{Bayesian Inference Library for \acs{CBC} \acs{GW} signal}
\acro{RAVEN}[RAVEN]{Rapid, on-source VOEvent Coincident Monitor}
\acro{MMA}[MMA]{Multi-Messenger Astronomy}
\acro{GWTC}[GWTC]{Gravitational-Wave Transient Catalog}
\acro{ML}[ML]{Machine Learning}
\acro{NN}[NN]{Neural Network}
\acro{QR}[QR]{Quantile Regression}
\acro{ROQ}[ROQ]{Reduced Order Quadrature}
\end{acronym}

\section{Introduction}
Over the course of three observing runs, the LIGO, Virgo,
and KAGRA (LVK)  have confirmed about $90$ \ac{CBC} detections of \acp{BNS}, \acp{BBH} and \acp{NSBH} \cite{gwtc1,gwtc2, gwtc3}. The \ac{O4} \footnote{\url{https://observing.docs.ligo.org/plan}} has seen more than 200 public alerts of \ac{GW} triggers \footnote{\url{https://gracedb.ligo.org/superevents/public/O4/}}. 

The majority of these alerts are generated from \ac{CBC} data analysis pipelines \citep{gstlal_becca,tito_pycbc,aubin_mbta,Chu_spiir} which are capable of detecting and analyzing triggers with latency less than 20 seconds \cite{sushant_llpaper}. The critical component of this process is matched filter \citep{findchirp,Cannon_earlywarning}. First, the detector data are filtered with a prebuilt template bank of waveforms \cite{shio_gstlal_templatebank}. Then the pipelines provide point estimates of the signal parameters by selecting the template with the highest \ac{SNR} or another suitable detection statistic  \cite{cody_gstlal,gstlal_tsukada,tito_pycbc,aubin_mbta,Chu_spiir}. 

Various injection studies show that systematic and statistical errors affect these point estimates \cite{gstlal_becca, deep_embright}. These uncertainties are especially relevant for high-mass \ac{BBH} systems because of the smaller number of templates in that region of the bank \cite{baird_degeneracy}. Additional uncertainties arise from the need to reduce the size of the template banks \cite{capano_alignedspins}. Therefore, the parameters of \ac{CBC} signals are not known in low latency with good accuracy; a full-scale \ac{PE} on time scales of hours to days is required to extract their best estimates. 

This situation is far from optimal. The physical parameters of \ac{BNS} and \ac{NSBH} mergers, in particular the chirp mass and total mass of the system, determine the disk wind and dynamic ejecta \cite{Margalit_2019}, and ultimately the post-merger evolution of the remnant. Therefore, estimating accurate parameters in low latency is critical to produce light curves and event sky localizations for \ac{EM} follow-up observations.

\ac{ML} methods \cite{deep_embright, ipam_embright, sushant_llpaper} have been successful in identifying source properties in real time, such as the presence of a \ac{NS}, the possibility of a post-merger remnant, the presence of mass-gap and sub-solar mass compact objects. However, obtaining accurate point estimates of \ac{CBC} parameters may prove challenging because of pipeline bias. In this paper, we take a different approach and develop a \ac{ML}-based method to provide real-time bounds on chirp mass, mass-ratio, and total mass parameters of a \ac{CBC} signal.  

We train a \ac{NN} \ac{QR} \cite{quantile_regression} algorithm on a dense dataset of simulated \ac{CBC} events. We derive bounds on chirp mass, mass-ratio, and total mass in the detector frame corresponding to selected quantile values. We find that the algorithm's accuracy is over 90\%. One of the key techniques that have been adopted to reduce the run time of online \ac{PE} is limiting the prior space of the sampling algorithm \cite{soichiro_bilby,rose_rapidpe, caitlin.thesis}. Our method could provide a dynamic way for doing so.

The paper is organized as follows. In Section \ref{sec:data_model} we describe the \ac{NN} architecture and the datasets for training and testing. In Section \ref{sec:O2_results} -- \ref{GWTC_results}, we present the results of the algorithm on various synthetic datasets and real events from the \ac{GWTC}. In section \ref{pe_analysis} we explore the possibility of using the \ac{NN} output to define priors for \ac{PE} in low latency. Conclusions are presented in Section \ref{sec:conclusions}.

\input{methods.tex}
\input{results}
\input{prior}
\input{conclusion}

\bibliography{biblography}
\input{appendix}
\end{document}

%% file: methods.tex
\section{Data and model architecture}\label{sec:data_model}

We consider two different datasets in our analysis. The first dataset has been utilized in the literature to train \ac{ML}-based algorithms for source property inference \cite{deep_embright,sushant_llpaper,ipam_embright}. It contains approximately $2.5 \times 10^5$ simulated \ac{CBC} signals injected in detector noise from the \ac{LV} \ac{O2} and filtered by the GstLAL search pipeline \cite{gstlal_becca} with a \ac{FAR} smaller than 1/month. The properties of this dataset are given in Table 1 of Ref.~\cite{deep_embright}.   We use this dataset for training, validation and testing purpose.

The second dataset is taken from the \ac{LVK} \ac{O3} replay \ac{MDC} \cite{sushant_llpaper}. It consists of simulated \ac{CBC} signals that were analyzed by online LVK search pipelines before the \ac{O4} in order to evaluate their performance. We utilize this dataset solely for testing purposes due to the limited $\mathcal O(10^3)$ number of simulated signals that the pipelines successfully recover.  We also utilize a subset of the \ac{O3} replay \ac{MDC} data for our \ac{PE} analysis in subsection \ref{pe_analysis}.

Both sets contain the intrinsic parameters of the simulated signals (detector frame masses and spins) and the template parameters that are recovered by the search pipelines:

\[
    \mathbf{x} = \left(m_1,~m_2,~ M_{\text{tot}},~ M_{\text{c}},~ q,~\chi_1^z,~ \chi_2^z,~ \text{SNR}\right)\,.
\]

As the chirp mass of the signal $M_{\rm c}$ increases, the error in its pipeline recovered value also increases \cite{gstlal_becca}. Therefore, following a binning approach adopted by the online PE algorithm \verb|bilby-pipe 1.3.0|, we split the data into four different bins according to the value of the chirp mass. Table \ref{table:bilbybins} lists the lower and upper bounds of these bins as well as their sizes. In order to train and test our method, we further split the datasets into ($72\% / 8\% /20\%$) train, validation. This produces a total of twelve sets.

\begin{table*}[htb]
    \centering
    \begin{tabular}{ccccc}
    \hline
 & \textbf{Lower Bound} & \textbf{Upper Bound} & \textbf{O2 dataset} & \textbf{O3 replay MDC} \\
    \hline \hline
    \textbf{BIN A} & $0 M_{\odot}$ & $1.465 M_{\odot}$ & 23187 & 1049  \\
    \textbf{BIN B} & $1.465 M_{\odot}$ & $2.234 M_{\odot}$ & 9374  & 1049 \\
    \textbf{BIN C} & $2.234 M_{\odot}$ & $12 M_{\odot}$ & 120937 & 1325 \\
    \textbf{BIN D} & $12 M_{\odot}$ & $\infty$ & 49909 & 2082\\
    \hline
    \end{tabular}
    \caption{Bins and corresponding number of events in the O2 and MDC datasets.}
    \label{table:bilbybins}
    \end{table*}

Figures \ref{fig:MDCvsO2_Mc}, \ref{fig:MDCvsO2_q}, and \ref{fig:MDCvsO2_Mtot} show the performance of the GStLAL pipeline on the O2 dataset (left panels) and the performance of the O4 pipelines on the replay MDC dataset (right panels) for different bins and different parameters. The maximum injected value for $M_{\rm c}$ and $M_{\rm tot}$ in the MDC dataset is higher than in the O2 dataset. The template bank for matched filtering in the MDC also covers a larger region of the parameter space compared to O2. As a consequence, the range of the recovered parameters is larger for the MDC dataset than the O2 dataset. Since we train solely on O2 data, we restrict the values of the MDC parameters within the O2 bounds. This reduces the number of injections for the MDC dataset listed in Table  \ref{table:bilbybins} to 594, 604, 668, and 262 injections for bins A, B, C, and D, respectively.

Our \ac{NN} model architecture is based on \ac{QR}. Regression is a supervised \ac{ML} technique that derives a relationship $y = f(\mathbf{x})$ between a response variable $y$ and its input vector $\mathbf{x}$. Conventional regression utilizes the least squares method to give a point estimate of the response variable as a conditional mean over $\mathbf{x}$. \ac{QR} is a variant of regression that estimates the conditional quantiles of the response variable. 

The advantage of QR over conventional regression is that it can provide conditional probability distributions rather than point estimates \cite{quantile_regression}. QR is also less affected by outliers than conventional methods since its loss function is based on absolute errors rather than squared errors.

\begin{figure*}[htb]
    \centering 
    \includegraphics[width=0.9\linewidth]{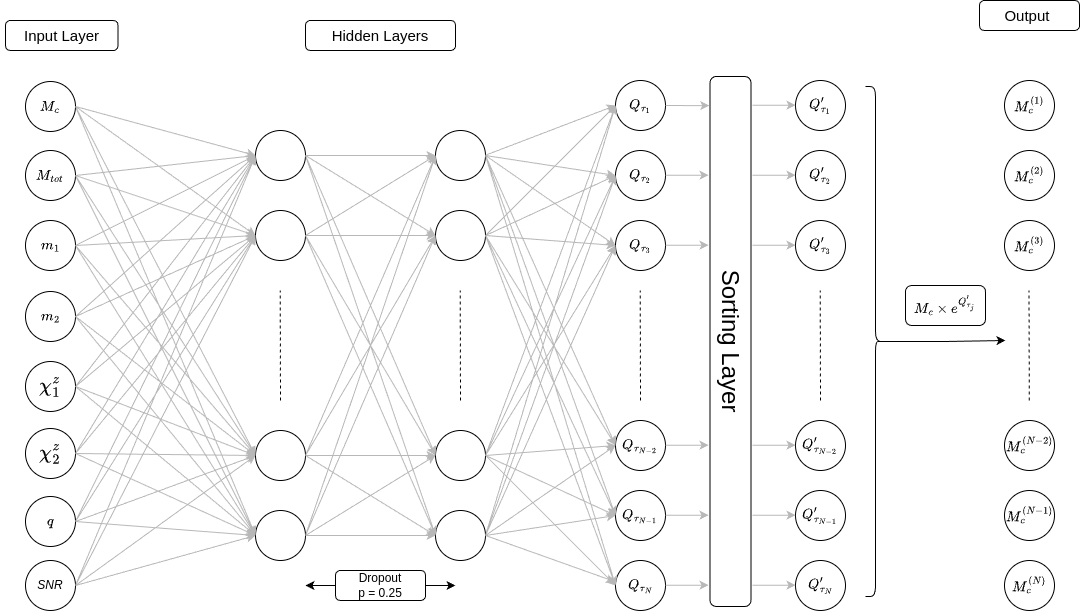}
    \caption{Illustration of the chirp mass \ac{NN} model. Two hidden layers of size $24 \times 12$ yield intermediate values $Q_{\tau_k}$ that are passed on to a sorting layer. The final output is produced by taking the product of $M_{\rm c}$ with the exponential of the sorted values. The models for $q$ and $M_{\rm tot}$ follow a similar structure, but with a different final layer.}
    \label{fig:network}
\end{figure*}

Typically, \ac{NN} implementations of \ac{QR} are more accurate
than conventional nonlinear \ac{QR} methods \cite{qrnn_xu}. The training of a \ac{QR} \ac{NN} model requires minimizing a quantile loss function. Given a random sample of response variables $y_i$, where $i= 1, \dots N$, the loss function for a quantile  $\tau \in (0, 1)$ is
\cite{cannon_qr_loss}
\begin{equation}
    \operatorname{QL} = \frac{1}{N}\sum_{i=1}^{N} \rho_\tau (y_i - y_\tau), \\
    \label{eq:quantileLoss} 
\end{equation}
where $y_\tau=\hbox{NN}(\mathbf{x};\tau)$ is the \ac{NN} estimate of the quantile and 
\begin{equation}
      \rho_\tau(u) = 
      \begin{cases}
        \tau u &\text{ if }  u \geq 0\,, \\
        (\tau-1) u &\text{ if } u < 0\,.
    \end{cases}
\end{equation}
In the case of a single \ac{NN} model predicting a number $T$ of quantiles $\tau_k$, where $k=1,\dots T$, the loss function can be generalized to 
\begin{equation}
    \operatorname{MQL} = \frac{1}{TN}\sum_{k=1}^{T}\sum_{i=1}^{N} \rho_{\tau_k} (y_i - y_{\tau_k})\,.
\label{eq:multi_quantile}
\end{equation}
In our study, the $y_i$ variables denote the CBC parameters for which we want to compute the quantiles ($M_c$, $M_{\rm tot}$, and $q$), with $i$ running over the number $N$ of samples in each batch. Our \ac{NN} \ac{QR} configuration is illustrated in Fig.~\ref{fig:network} and described below following the PyTorch \cite{pytorch2024} library nomenclature \cite{pytorch}. The architecture consists of a fully connected \ac{NN} with two (24, 12) hidden layers. We choose \texttt{LeakyReLU} \cite{leaky_relu} as the activation function for each hidden layer and between each of them we insert a \texttt{dropout} layer to randomly zero out 25\% of the nodes in the training process. We normalize and rescale the input before passing it to the network with a \texttt{LayerNorm} layer. The output of the hidden layers is sorted in ascending order by a \verb|SoftSort| \cite{pmlr-v119-blondel20a}. This configuration prevents possible inversions of quantile values in the output. The latter is a set of unconstrained real numbers that is conditioned according to the specific training parameter: chirp mass, ${M_{\rm c}}$, mass ratio, $q$, and total component mass $M_{\rm tot}$. For the mass ratio we apply a sigmoid function to constrain the output of the sorting layer to the physical range $(0,1)$. In the case of chirp mass and total component mass, we take the exponential of the sorting layer output and multiply it by the corresponding recovered parameters. For example, in the case of the chirp mass we have $M_{\rm c}(\tau_i, \mathbf{x}) =  M_{\rm c}\exp (Q^\prime_{\tau_i})$, where $Q^\prime_{\tau_i}$ indicates the output of the sorted layer. 

We train the model to produce $T=37$ different quantiles with the stochastic gradient descent and loop through the whole dataset 100 times (epochs), dividing it into mini-batches of $N=400$ samples. The values of the quantiles are selected such that they provide a denser representation at the lower and upper bounds of the quantile range.
We utilize the \verb|AdamW| \cite{adam} optimizer with 
the \texttt{CosineAnnealingWarmRestarts} learning rate scheduler. This ensures that the learning rate, which is related to the magnitude of the gradient descent steps, decreases monotonically across the epochs as a cosine curve. The model weights are selected across epochs to minimize the loss on the validation set.
We select the training hyperparameters by testing multiple configurations of the optimizer, learning rate, and the number of nodes in the hidden layers. For the sake of streamlining the training process, we settle on a common choice of hyperparameters that is unique through all bins while keeping the validation loss to a minimum. The initial learning rates are set to $5\times 10^{-3}$ ($5 \times 10^{-2}$) for $M_{\rm c}$ and $M_{\rm tot}$ ($q$).

\begin{figure*}[htb]
    \centering 
    \includegraphics[width=1\linewidth]{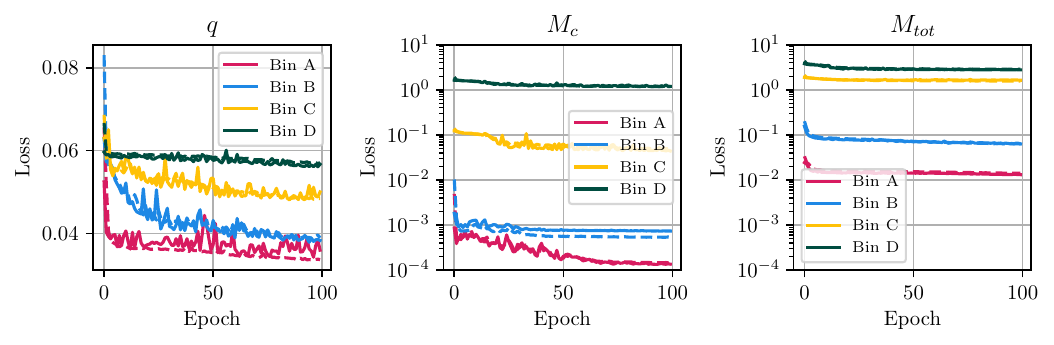}
    \caption{Training (dashed lines) and validation (solid lines) loss curves for mass ratio, chirp mass, and total mass.}
    \label{fig:loss_curve}
\end{figure*}

Figure \ref{fig:loss_curve} displays the loss curves for the twelve independent models (four chirp mass bins $\times$ three parameters). As the number of the training epoch increases, both training and validation losses decrease, becoming essentially constant after $\sim$ 50 epochs. 
The validation loss follows the training loss in all models, indicating that there is no overfitting.

To assess the performance of our models we define two metrics, accuracy and interval width. For each quantile with target accuracy $\hbox{TA} =\tau_{T-k} - \tau_k$, we build the confidence interval $I_k = \left[X^{(k)}, X^{(T-k)}\right]$, where $X$ is any of the training parameters. If the true (injected) value of the parameter lies within the interval, we call $I_k$ \textit{accurate}. The fraction of accurate intervals across the dataset defines the accuracy of the models. The interval width is defined as $X^{(T-k)} - X^{(k)}$. An optimal model produces intervals with maximum accuracy and smallest width compatible with the uncertainties of the recovered parameter.

%% file: results.tex
%PLAN: 
% widths are in line with pipeline-relative errors. (histogram plots) 
% - target accuracy is okay on O2 - okayish on O3 but motivated by differences in dataset. Have to say this? 
% - summary table with accuracies, write accuracies of model used in 102 events runs
% - accuracy plots: collapse them? 
% - test on bilby: gain in likelihood evaluations , pp-plots and cumulative searched area plots. 

%\section{Results}

\section{O2 results}\label{sec:O2_results}
\begin{figure}[h]
    \centering
    \includegraphics[width=1\linewidth]{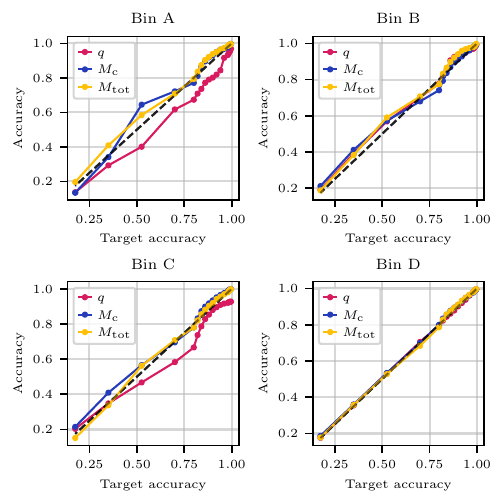}
    \caption{O2 testing dataset accuracy as a function of target accuracy for the three \ac{NN} models and different bins. The black dashed lines indicate ideal performance.}
    \label{fig:O2-pp}
\end{figure}

Figure \ref{fig:O2-pp} shows the performance of the models on the O2 testing dataset. The dashed diagonal lines indicate ideal performance. The model for the mass ratio ($q$) slightly underperforms for the A and C bins.  We pass the final network output for this model onto a sigmoid function to constrain the predictions between $q=0$ and $q=1$. However, since the sigmoid function only reaches these boundaries asymptotically, the upper bound of the predicted interval is never reached. This leads to a slight bias for the events with $q \sim 1$.  

\begin{table}[h!]
    \centering
    \begin{tabular}{cccc}
        \hline
        \textbf{Bin} ~&~ \textbf{Parameter} ~&~ \textbf{Accuracy} ~&~ \textbf{Median width} \\
        \hline
         ~&~ $q$ ~&~ 0.916  ~&~ ~0.512\\
        A ~&~ $M_{\rm c}$ ~&~ 0.968  ~&~ ~0.002 \\
        ~&~ $M_{\rm tot}$ ~&~ 0.974 ~&~ ~0.234 \\
        \hline
         ~&~ $q$ ~&~ 0.962 ~&~ ~0.642 \\
       B ~&~ $M_{\rm c}$ ~&~ 0.961 ~&~ ~0.007 \\
        ~&~ $M_{\rm tot}$ ~&~ 0.975 ~&~ ~0.977 \\
        \hline
         ~&~ $q$ ~&~ 0.916 ~&~ ~0.893 \\
       C ~&~ $M_{\rm c}$ ~&~ 0.976 ~&~ ~0.159 \\
        ~&~ $M_{\rm tot}$ ~&~ 0.970 ~&~ 17.217 \\
        \hline
         ~&~ $q$ ~&~ 0.961 ~&~ ~0.731 \\
        D ~&~ $M_{\rm c}$ ~&~ 0.963 ~&~ 17.232 \\
        ~&~ $M_{\rm tot}$ ~&~ 0.966 ~&~ 40.789 \\
        \hline
    \end{tabular}
    \caption{Accuracy and median width of the intervals corresponding to a target accuracy of $96\%$. The values for the chirp mass and total mass are in solar mass units ($M_\odot$).}
    \label{tab:model_results_O2}
\end{table}

\begin{figure}
    \centering
\includegraphics[width=\linewidth]{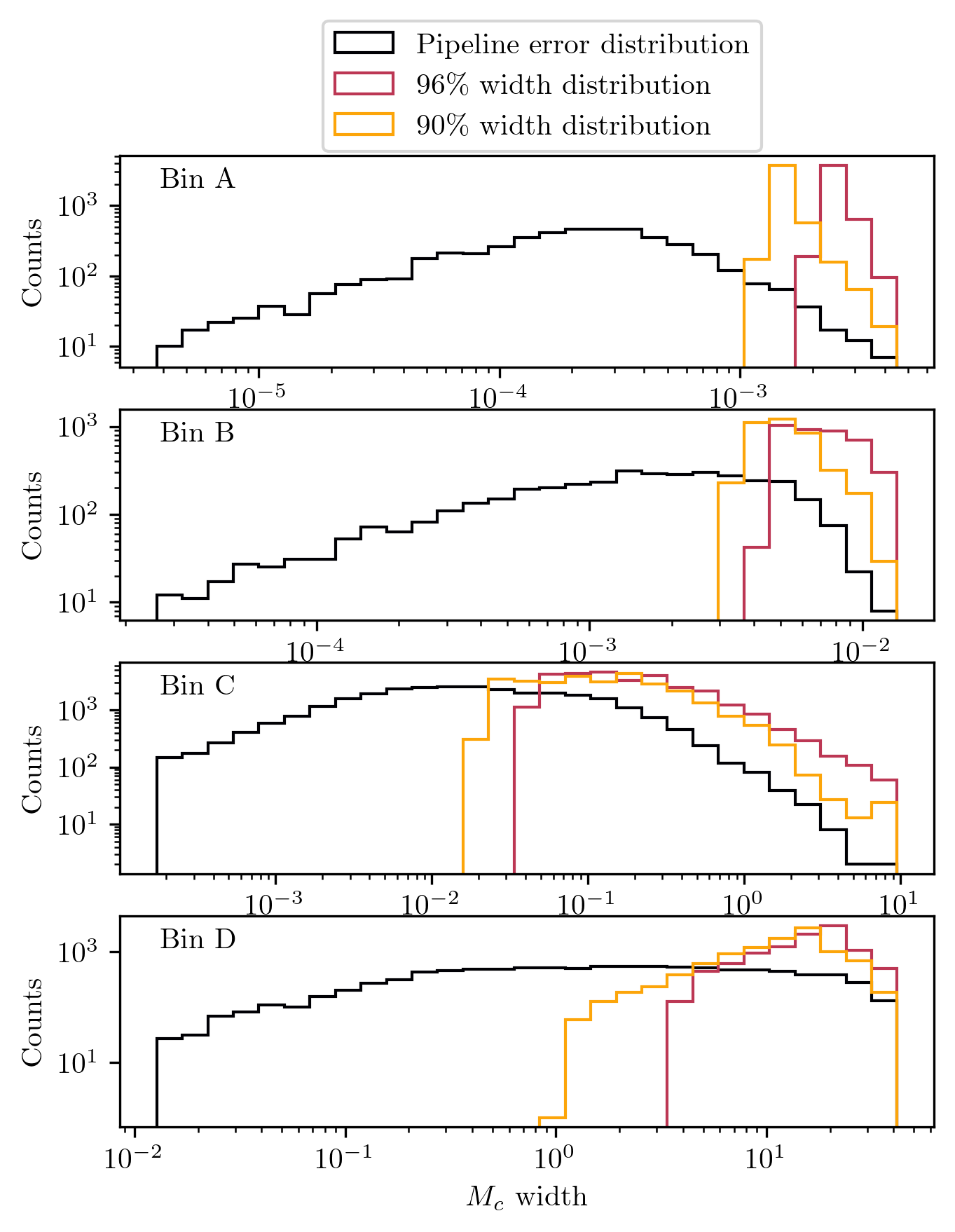}
    \caption{The figure shows distribution of pipeline error width (defined as difference between injected and recovered values) and  \ac{NN} widths for 96\% and 90\% target accuracies across different bins in O2 testing dataset. The \ac{NN} widths across all bins are comparable to the pipeline maximum errors and match the tail of the distributions. Reducing the target accuracy from 96\% to 90\% causes the NN widths distributions to shift towards the left because of the smaller widths.}
    \label{fig:relative_errs}
\end{figure}

Table \ref{tab:model_results_O2} lists the accuracy and interval median width for different \ac{NN} models and bins. The accuracy is estimated on the testing sets after setting a 96\% target accuracy on the training set. The accuracy is consistently above 90\% for all models and bins. 
Figure \ref{fig:relative_errs} shows that the \ac{NN} interval widths are typically larger than the pipeline errors with their largest values being comparable. Reducing the target accuracy leads to smaller width distributions.

\section{O3 MDC results}\label{sec:O3_results}

\begin{figure}[h]
    \centering
    \includegraphics[width=1\linewidth]{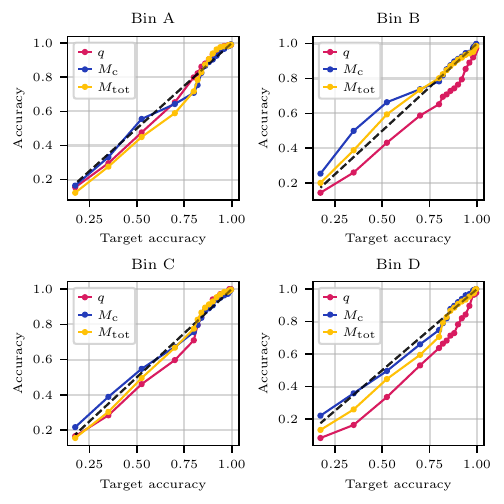}
    \caption{O3 Replay MDC testing dataset accuracy as a function of target accuracy for the three \ac{NN} models and different bins. The black dashed lines indicate ideal performance.}
    \label{fig:MDC-pp}
\end{figure}

\begin{table}[h!]
    \centering
    \begin{tabular}{cccc}
        \hline
        \textbf{Bin} ~&~ \textbf{Parameter} ~&~ \textbf{Accuracy} ~&~ \textbf{Median width} \\
        \hline
         ~&~ $q$ ~&~ 0.983 ~&~ ~0.522 \\
        A ~&~ $M_{\rm c}$ ~&~ 0.966 ~&~ ~0.003 \\
        ~&~ $M_{\rm tot}$ ~&~ 0.981 ~&~ ~0.284 \\
        \hline
         ~&~ $q$ ~&~ 0.891 ~&~ ~0.643 \\
       B ~&~ $M_{\rm c}$ ~&~ 0.947 ~&~ ~0.005\\
        ~&~ $M_{\rm tot}$ ~&~ 0.944 ~&~ ~0.832 \\
        \hline
         ~&~ $q$  ~&~ 0.978 ~&~ ~0.918 \\
       C ~&~ $M_{\rm c}$ ~&~ 0.964 ~&~  ~0.145 \\
        ~&~ $M_{\rm tot}$ ~&~ 0.984 ~&~ 14.657 \\
        \hline
         ~&~ $q$  ~&~ 0.897 ~&~ ~0.825 \\
        D ~&~ $M_{\rm c}$  ~&~ 0.969 ~&~ 19.83 \\
        ~&~ $M_{\rm tot}$ ~&~ 0.954 ~&~ 70.291 \\
        \hline
    \end{tabular}
    \caption{Accuracy and median width of the intervals corresponding to a target accuracy of $96\%$. The values for the chirp mass and total mass are in solar mass units ($M_\odot$).}
    \label{tab:model_results_MDC}
\end{table}

Figure \ref{fig:MDC-pp} and Table \ref{tab:model_results_MDC} show the results of the \ac{NN} models trained on O2 data on the O3 MDC set. Since the parameter space of the O2 and the O3 MDC datasets are different, we evaluate the models on a reduced O3 dataset (see Sec.~\ref{sec:data_model}.) The different trends of the parameters in the bins with respect to O2 are likely due to the inclusion of multiple pipelines in the O3 dataset, while training on the output of a single pipeline (GstLAL) in the O2 data. Changes in pipelines and their template banks leading up to \ac{O4} also affect injection recovery across different datasets. In spite of this, Fig.~\ref{fig:MDC-pp} and table \ref{tab:model_results_MDC} confirm that the \ac{NN} models perform similarly on the O3 set as they do on O2.

\begin{figure}[]
    \centering
    \includegraphics[width=0.97\linewidth]{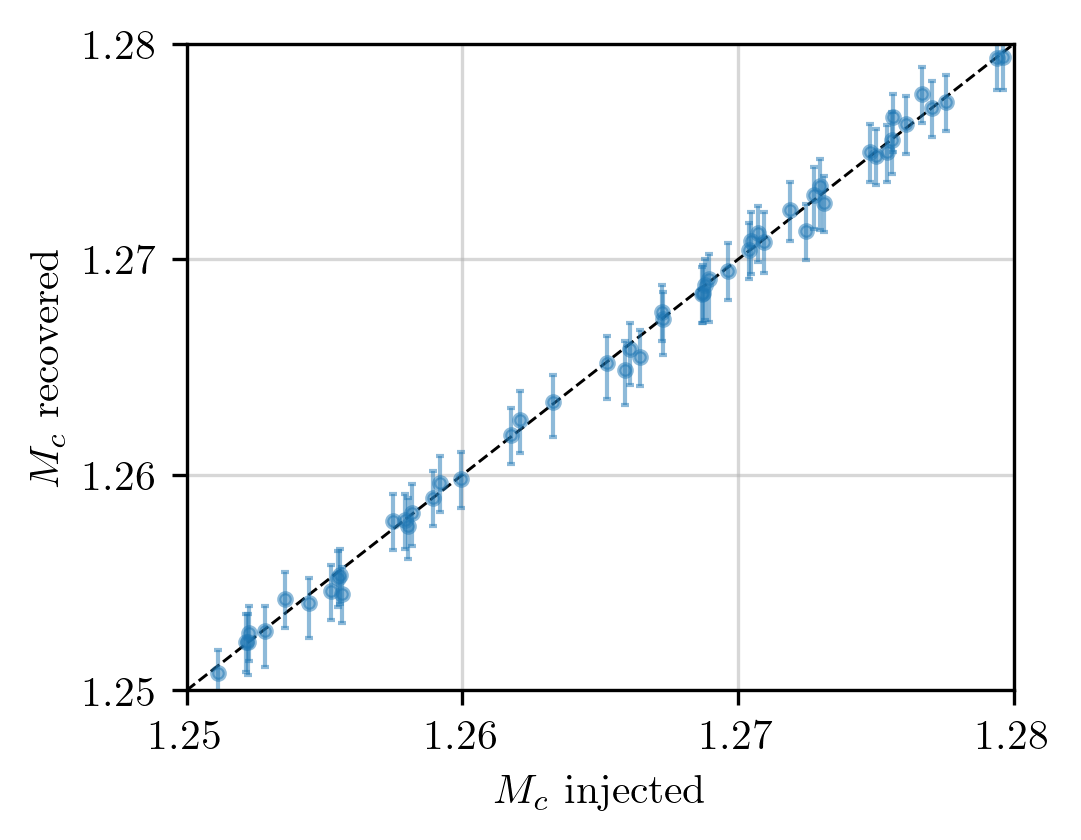}
    \includegraphics[width=0.97\linewidth]{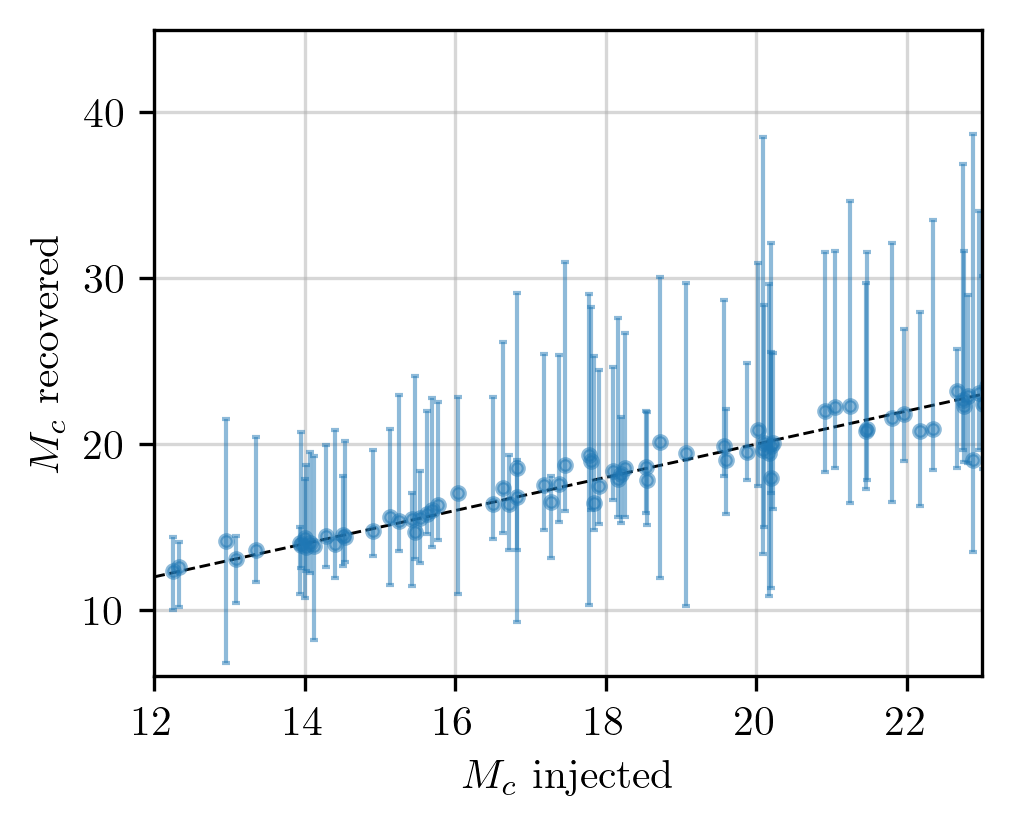}
    \includegraphics[width=0.97\linewidth]{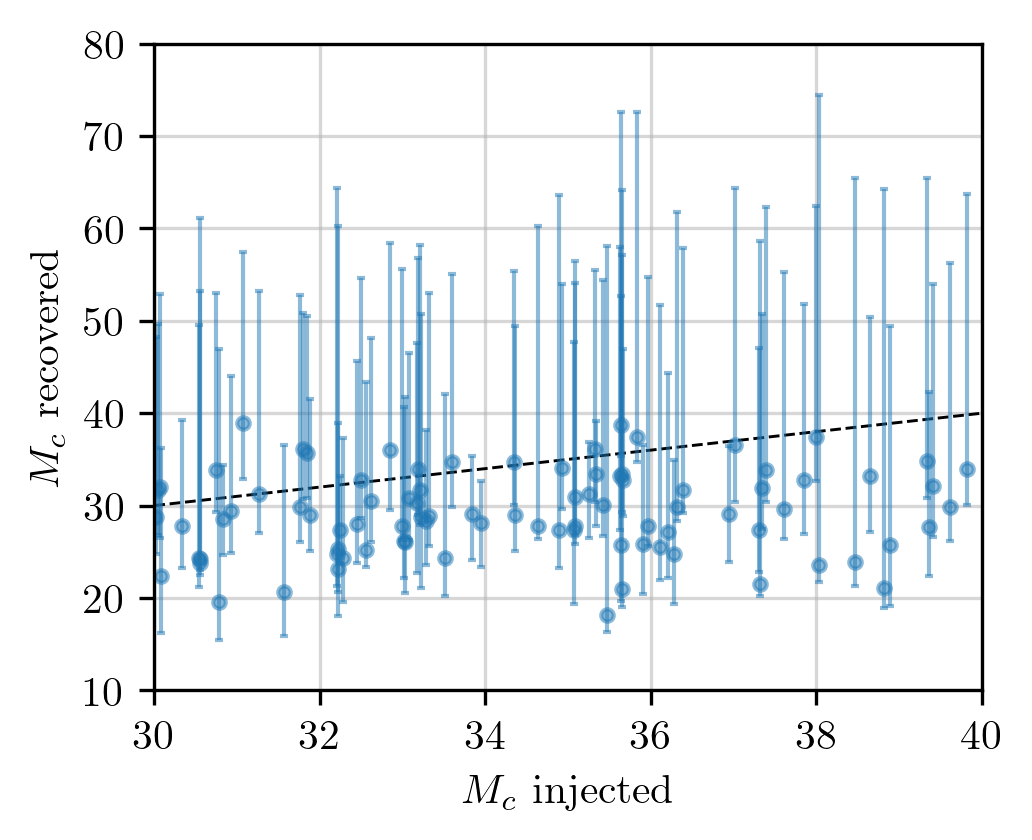}
    \caption{Pipeline recovered chirp mass and \ac{NN} predicted intervals vs.\ injected chirp mass for the O3 MDC dataset. The dots represent the pipeline point estimate. The  error bars represent the intervals predicted by the \ac{NN} model. The black dashed lines represent perfect injection recovery. Intervals are estimated with 96\% target accuracy. The three panels show the NN estimates for different mass ranges.}
    \label{fig:MDC_chirp}
\end{figure}

Figure \ref{fig:MDC_chirp} illustrates the output of the chirp mass model for 96\% target accuracy across three random injected chirp mass ranges. The three panels show the pipeline recovered chirp mass and the \ac{NN} predicted intervals vs.\ the injected chirp mass for a sample of three different $M_{\rm c}$ ranges. The pipelines recover the chirp mass fairly accurately between $M_{\rm c} \in$ [1.25, 1.28]$M_\odot$ (top panel). Therefore, the \ac{NN} predicted intervals are of the order of $10^{-3}~M_\odot$. Even when the injected chirp mass is not accurately recovered by the pipelines, the \ac{NN} intervals still encompass the true injected value. As the injected $M_{\rm c}$ increases, the pipeline accuracy decreases (middle panel). The median interval width also increase, ensuring that the injected $M_{\rm c}$ remains included within the \ac{NN} prediction. In the $M_{\rm c} \in$ [30, 40]$M_\odot$ range (bottom panel), the pipelines underestimate most of the injected chirp masses. The \ac{NN} model compensates for this bias by dynamically extending the interval upper bound. Overall, the \ac{NN} provides accurate intervals for the O3 Replay MDC even if it was trained on O2 data.

\section{Performance on a synthetic dataset}\label{sec:synthetic_results}

\begin{figure}[h!]
    \centering
    \includegraphics[width=1\linewidth]{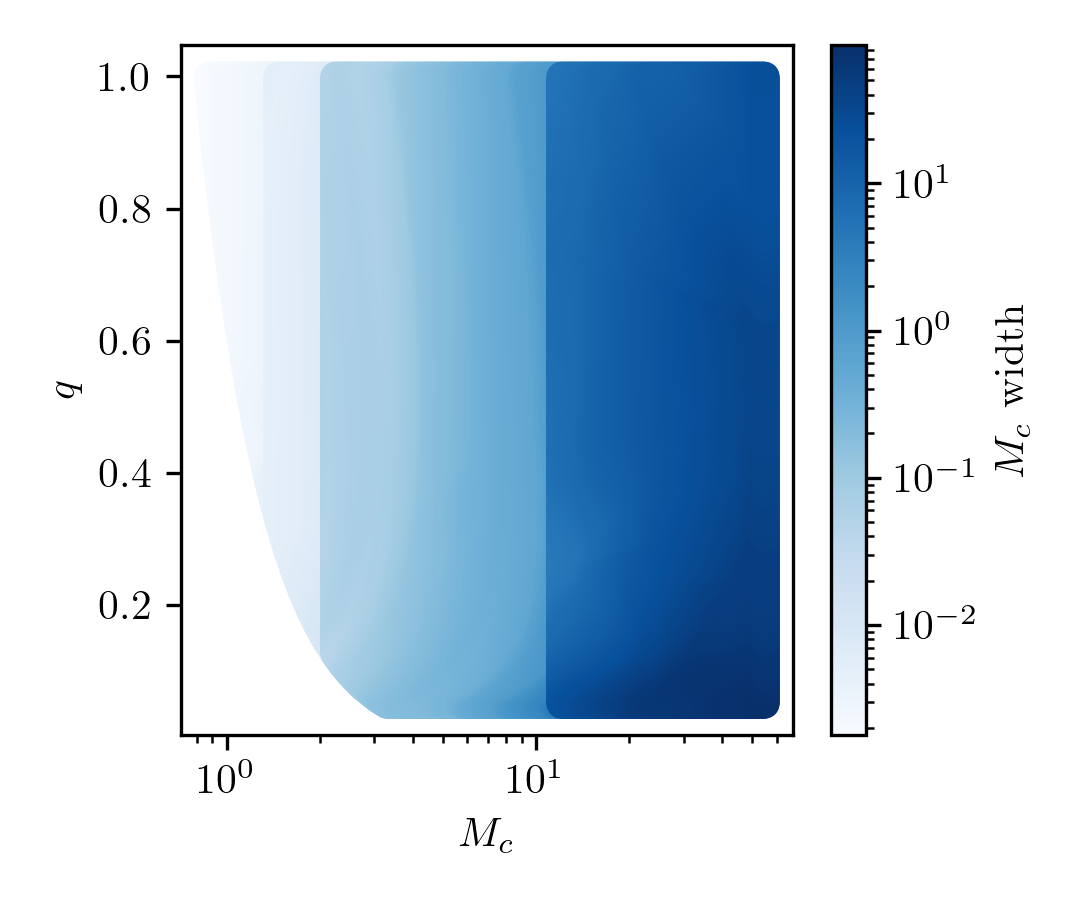}
    \caption{Width of the chirp mass intervals predicted
            by the neural networks with 96\% target accuracy, 
            as function of the input mass parameters on the 
            synthetic dataset. The SNR is fixed to 15 and the
            input spins are set $\chi_1^z, \chi_2^z =0$.   The discontinuity in the
            colors are due to different models being used in
            different regions.}
    \label{fig:synthetic_width}
\end{figure}

\begin{figure}[h!]
    \centering
    \includegraphics[width=1\linewidth]{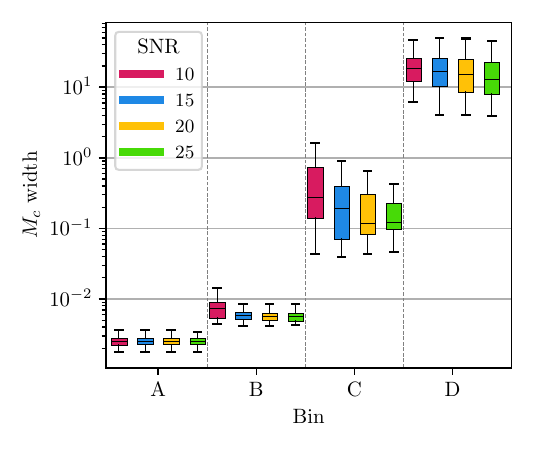}
    \caption{Interval widths for the 
            synthetic dataset and different bins for various \ac{SNR} values. The edges of the boxes indicate the 1st and 3rd quartile. The whiskers indicate the minimum and maximum value of the interval widths. The black lines indicate median values.}
    \label{fig:synthetic_width_snr}
\end{figure}

In order to assess the \ac{NN} performance on the full mass parameter space, we create a synthetic dataset. We choose a log-uniform distribution for $M_{\rm c}$ between 0.5 and 55 $M_\odot$ to avoid oversampling high values of the chirp mass and a uniform distribution for $q$ between 0.05 and 1. As we are mainly interested in assessing the performance on the chirp mass and the mass ratio, we set the $z-$spin components $\chi_1^z, \chi_2^z$ to zero and fix the \ac{SNR} to 10, 15, 20, and 25. Figure \ref{fig:synthetic_width} shows the \ac{NN} interval width with 96\% target accuracy. Overall, in each bin the interval width decreases as $q$ increases. The width also generally increases with $M_{\rm c}$ at fixed $q$. These trends can also be seen in Fig.~\ref{fig:synthetic_width_snr} which shows the widths of the chirp mass intervals for different values of the \ac{SNR}. We choose the \ac{SNR} metric to assess the performance of the \ac{NN} because it is a common statistic across all pipelines with a clear definition that does not depend on the each pipeline implementations. The median widths decrease with increasing \ac{SNR} and  other input parameters held fixed. This is more evident for bins B, C, and D.


\section{Performance on GWTC events}\label{GWTC_results}
We further assess the performance of the \ac{NN} models on the events from the \ac{GWTC} \cite{gwtc2_1, gwtc3} that passed the low-latency alert threshold. We choose the input parameters for the \ac{NN} from low-latency templates with highest \ac{SNR}. The \ac{GWTC} catalog provides median values and 90\% symmetric credible intervals for the source parameters. The $M_{\rm c}^{\rm PE}$, $q^{\rm PE}$, and $M_{\rm tot}^{\rm PE}$ columns in Table \ref{tab:real_events} list median detector frame values for these parameters obtained from the \ac{GWTC} posterior distributions. We compare these values to the width intervals with 96\% target accuracy.

The \ac{NN} intervals of all events include their corresponding \ac{GWTC} median values with the exception of GW190521\_030229. This event is reported in \ac{GWTC} with a chirp mass of $\sim 101 M_\odot$, the largest in the catalog. However, its low-latency chirp mass estimate is $M_{\rm c}=32.96~M_\odot$. Moreover, the training dataset is restricted to events with chirp mass values below $60 M_\odot$. Therefore, it is not surprising that the \ac{NN} model underestimates the actual value of the event.

\begin{center}
\begin{table*}[tp]
    \centering
    \renewcommand{\arraystretch}{1.18}
    % \rowcolors{2}{gray!25}{white}
    \setlength{\tabcolsep}{0pt}
    \begin{tabular}{lcrrrrrr}
     \hline
        \textbf{Event\_id} ~&~ \textbf{GWTC} ~&~ \boldmath$M_{\rm c}^{\rm PE}$ ~&~ \boldmath$M_{\rm c}$ \textbf{Interval} ~&~ \boldmath$q^{\rm PE}$ ~&~ \boldmath$q$ \textbf{Interval} ~&~ \boldmath$M_{\rm tot}^{\rm PE}$ ~&~ \boldmath$M_{\rm tot}$ \textbf{Interval} \\
        \hline
        GW190408\_181802 ~&~ 2.1 ~&~ $23.8_{-1.4}^{+1.2}        $ ~&~ [20.7, 35.0]      ~&~  $0.75_{-0.25}^{+0.21}$  ~&~ [0.27, 0.98]  ~&~  $55.8_{-3.2}^{+3.1}    $  ~&~ [49.7, 78.2] \\ \hline
        GW190412         ~&~ 2.1 ~&~ $15.24_{-0.2}^{+0.3}       $ ~&~ [14.1, 21.0]      ~&~  $0.32_{-0.09}^{+0.17}$  ~&~ [0.25, 0.98]  ~&~  $42.1_{-4.5}^{+5.3}    $  ~&~ [32.6, 48.2] \\ \hline
        GW190421\_213856 ~&~ 2.1 ~&~ $45.9_{-6.2}^{+5.6}        $ ~&~ [34.3, 66.5]      ~&~  $0.78_{-0.32}^{+0.19}$  ~&~ [0.27, 0.98]  ~&~  $107.3_{-11.6}^{+12.4} $  ~&~ [88.9, 152.3] \\ \hline
        GW190425         ~&~ 2.1 ~&~ $1.4868_{-0.0003}^{+0.0003}$ ~&~ [1.4847, 1.4893]  ~&~  $0.89_{-0.14}^{+0.09}$  ~&~ [0.34, 0.99]  ~&~  $3.422_{-0.006}^{+0.03}$  ~&~ [3.240, 3.944] \\ \hline
        GW190503\_185404 ~&~ 2.1 ~&~ $37.8_{-6.7}^{+6.0}        $ ~&~ [15.9, 44.0]      ~&~  $0.69_{-0.29}^{+0.27}$  ~&~ [0.12, 0.98]  ~&~  $89.6_{-13.0}^{+13.2}  $  ~&~ [46.3, 116.6] \\ \hline
        GW190512\_180714 ~&~ 2.1 ~&~ $18.5_{-0.6}^{+0.6}        $ ~&~ [15.3, 27.1]      ~&~  $0.53_{-0.18}^{+0.36}$  ~&~ [0.25, 0.98]  ~&~  $45.2_{-2.7}^{+4.1}    $  ~&~ [38.2, 62.3] \\ \hline
        GW190513\_205428 ~&~ 2.1 ~&~ $30.4_{-3.2}^{+5.8}        $ ~&~ [24.0, 42.8]      ~&~  $0.50_{-0.19}^{+0.41}$  ~&~ [0.25, 0.98]  ~&~  $75.4_{-8.1}^{+14.1}   $  ~&~ [59.5, 102.0] \\ \hline
        GW190517\_055101 ~&~ 2.1 ~&~ $35.7_{-5.5}^{+4.2}        $ ~&~ [30.0, 46.6]      ~&~  $0.61_{-0.30}^{+0.32}$  ~&~ [0.35, 0.98]  ~&~  $85.7_{-8.5}^{+9.9}    $  ~&~ [73.7, 114.0] \\ \hline
        GW190519\_153544 ~&~ 2.1 ~&~ $65.0_{-10.8}^{+8.6}       $ ~&~ [45.2, 78.4]      ~&~  $0.62_{-0.22}^{+0.26}$  ~&~ [0.27, 0.97]  ~&~  $154.8_{-18.1}^{1+8.6} $  ~&~ [113.2, 190.3] \\ \hline
        GW190521\_030229 ~&~ 2.1 ~&~ $101.0_{-33.7}^{+28.9}     $ ~&~ [27.2, 88.8]      ~&~  $0.58_{-0.37}^{+0.33}$  ~&~ [0.09, 0.98]  ~&~  $243.2_{-36.0}^{+58.2} $  ~&~ [67.9, 255.1] \\ \hline
        GW190521\_074359 ~&~ 2.1 ~&~ $39.6_{-2.5}^{+3.0}        $ ~&~ [34.3, 58.0]      ~&~  $0.77_{-0.20}^{+0.18}$  ~&~ [0.26, 0.98]  ~&~  $92.3_{-5.0}^{+6.8}    $  ~&~ [80.4, 126.3] \\ \hline
        GW190602\_175927 ~&~ 2.1 ~&~ $72.7_{-18.2}^{+12.2}      $ ~&~ [27.7, 82.0]      ~&~  $0.62_{-0.34}^{+0.32}$  ~&~ [0.18, 0.98]  ~&~  $173.8_{-24.9}^{+25.6} $  ~&~ [81.4, 211.3] \\ \hline
        GW190630\_185205 ~&~ 2.1 ~&~ $29.4_{-1.6}^{+2.0}        $ ~&~ [23.8, 42.9]      ~&~  $0.68_{-0.21}^{+0.27}$  ~&~ [0.25, 0.98]  ~&~  $69.7_{-3.5}^{+4.7}    $  ~&~ [60.0, 100.8] \\ \hline
        GW190701\_203306 ~&~ 2.1 ~&~ $55.5_{-8.0}^{+7.3}        $ ~&~ [35.1, 72.9]      ~&~  $0.76_{-0.31}^{+0.21}$  ~&~ [0.28, 0.98]  ~&~  $130.2_{-15.2}^{+16.0} $  ~&~ [93.9, 173.8] \\ \hline
        GW190706\_222641 ~&~ 2.1 ~&~ $74.7_{-18.9}^{+13.4}      $ ~&~ [60.4, 98.2]      ~&~  $0.53_{-0.24}^{+0.34}$  ~&~ [0.33, 0.99]  ~&~  $182.6_{-27.9}^{+25.9} $  ~&~ [142.6, 221.1] \\ \hline
        GW190707\_093326 ~&~ 2.1 ~&~ $9.89_{-0.09}^{+0.10}      $ ~&~ [9.72, 10.66]     ~&~  $0.66_{-0.20}^{+0.27}$  ~&~ [0.01, 0.99]  ~&~  $23.3_{-0.6}^{+1.5}    $  ~&~ [19.1, 62.4] \\ \hline
        GW190720\_000836 ~&~ 2.1 ~&~ $10.36_{-0.10}^{+0.10}     $ ~&~ [10.16, 11.15]    ~&~  $0.52_{-0.23}^{+0.35}$  ~&~ [0.00, 0.99]  ~&~  $25.3_{-1.5}^{+4.2}    $  ~&~ [18.7, 71.3] \\ \hline
        GW190727\_060333 ~&~ 2.1 ~&~ $44.8 _{-5.7}^{+5.3}       $ ~&~ [28.5, 52.4]      ~&~  $0.79_{-0.30}^{+0.18}$  ~&~ [0.31, 0.98]  ~&~  $104.6_{-10.7}^{+12.7} $  ~&~ [71.3, 122.8] \\ \hline
        GW190728\_064510 ~&~ 2.1 ~&~ $10.13_{-0.08}^{+0.10}     $ ~&~ [9.79, 10.29]     ~&~  $0.64_{-0.35}^{+0.30}$  ~&~ [0.01, 0.98]  ~&~  $23.9_{-0.7}^{+5.2}    $  ~&~ [20.2, 65.0] \\ \hline
        GW190814         ~&~ 2.1 ~&~ $6.41 _{-0.01}^{+0.01}     $ ~&~ [6.29, 6.52]      ~&~  $0.11_{-0.01}^{+0.01}$  ~&~ [0.05, 0.99]  ~&~  $27.2_{-1.3}^{+1.4}    $  ~&~ [15.2, 40.0] \\ \hline
        GW190828\_063405 ~&~ 2.1 ~&~ $33.8_{-2.5}^{+2.6}        $ ~&~ [27.3, 49.4]      ~&~  $0.82_{-0.24}^{+0.15}$  ~&~ [0.26, 0.98]  ~&~  $78.5_{-5.3}^{+6.1}    $  ~&~ [67.9, 114.2] \\ \hline
        GW190828\_065509 ~&~ 2.1 ~&~ $17.3_{-0.7}^{+0.6}        $ ~&~ [14.6, 26.0]      ~&~  $0.44_{-0.16}^{+0.38}$  ~&~ [0.24, 0.98]  ~&~  $43.9_{-3.7}^{+5.7}    $  ~&~ [36.1, 63.3] \\ \hline
        GW190915\_235702 ~&~ 2.1 ~&~ $32.3_{-3.1}^{+3.0}        $ ~&~ [24.0, 36.9]      ~&~  $0.76_{-0.29}^{+0.20}$  ~&~ [0.31, 0.98]  ~&~  $75.8_{-6.7}^{+7.6}    $  ~&~ [58.1, 93.6] \\ \hline
        GW190924\_021846 ~&~ 2.1 ~&~ $6.43_{-0.02}^{+0.03}      $ ~&~ [6.30, 6.60]      ~&~  $0.58_{-0.30}^{+0.32}$  ~&~ [0.04, 0.96]  ~&~  $15.4_{-0.6}^{+3.2}    $  ~&~ [12.4, 44.0] \\ \hline
        GW190930\_133541 ~&~ 2.1 ~&~ $9.85_{-0.22}^{+0.15}      $ ~&~ [9.58, 10.11]     ~&~  $0.48_{-0.26}^{+0.42}$  ~&~ [0.02, 0.97]  ~&~  $24.3_{-1.8}^{+6.8}    $  ~&~ [19.0, 77.7] \\ \hline
        GW191105\_143521 ~&~ 3   ~&~ $9.57_{-0.13}^{+0.11}      $ ~&~ [9.37, 10.21]     ~&~  $0.72_{-0.31}^{+0.24}$  ~&~ [0.00, 0.99]  ~&~  $22.3_{-0.5}^{+2.3}    $  ~&~ [18.6, 68.5] \\ \hline
        GW191109\_010717 ~&~ 3   ~&~ $60.1_{-9.3}^{+9.8}        $ ~&~ [44.7, 80.0]      ~&~  $0.73_{-0.23}^{+0.21}$  ~&~ [0.33, 0.98]  ~&~  $140.4_{-16.7}^{+21.3} $  ~&~ [110.9, 189.6] \\ \hline
        GW191129\_134029 ~&~ 3   ~&~ $8.48_{-0.05}^{+0.05}      $ ~&~ [8.33, 8.77]      ~&~  $0.63_{-0.29}^{+0.31}$  ~&~ [0.01, 0.98]  ~&~  $20.1_{-0.6}^{+2.9}    $  ~&~ [15.0, 48.9] \\ \hline
        GW191204\_171526 ~&~ 3   ~&~ $9.69_{-0.05}^{+0.05}      $ ~&~ [9.53, 10.01]     ~&~  $0.69_{-0.25}^{+0.25}$  ~&~ [0.02, 0.97]  ~&~  $22.7_{-0.4}^{+1.9}    $  ~&~ [17.8, 55.5] \\ \hline
        GW191215\_223052 ~&~ 3   ~&~ $24.8_{-1.4}^{+1.4}        $ ~&~ [21.3, 37.7]      ~&~  $0.73_{-0.27}^{+0.23}$  ~&~ [0.21, 0.98]  ~&~  $58.4_{-3.6}^{+4.8}    $  ~&~ [53.1, 92.3] \\ \hline
        GW191216\_213338 ~&~ 3   ~&~ $8.93_{-0.04}^{+0.05}      $ ~&~ [8.81, 9.30]      ~&~  $0.63_{-0.28}^{+0.30}$  ~&~ [0.04, 0.97]  ~&~  $21.1_{-0.6}^{+2.9}    $  ~&~ [17.8, 63.2] \\ \hline
        GW191222\_033537 ~&~ 3   ~&~ $51.0_{-6.4}^{+7.1}        $ ~&~ [23.3, 64.2]      ~&~  $0.79_{-0.31}^{+0.18}$  ~&~ [0.18, 0.98]  ~&~  $119.2_{-13.0}^{+15.7} $  ~&~ [66.7, 167.0] \\ \hline
        GW200105\_162426 ~&~ 3   ~&~ $3.619_{-0.008}^{+0.011}   $ ~&~ [3.607, 3.761]    ~&~  $0.22_{-0.07}^{+0.15}$  ~&~ [0.01, 0.99]  ~&~  $11.3_{-1.8}^{+2.2}    $  ~&~ [8.7, 27.8] \\ \hline
        GW200112\_155838 ~&~ 3   ~&~ $33.9_{-2.3}^{+2.9}        $ ~&~ [30.4, 50.7]      ~&~  $0.80_{-0.25}^{+0.17}$  ~&~ [0.26, 0.98]  ~&~  $79.0_{-5.0}^{+6.4}    $  ~&~ [73.5, 110.9] \\ \hline
        GW200115\_042309 ~&~ 3   ~&~ $2.582_{-0.005}^{+0.006}   $ ~&~ [2.542, 2.646]    ~&~  $0.19_{-0.06}^{+0.12}$  ~&~ [0.11, 0.51]  ~&~  $8.5_{-1.4}^{+1.5}     $  ~&~ [5.1, 10.5] \\ \hline
        GW200128\_022011 ~&~ 3   ~&~ $49.8_{-6.5}^{+7.2}        $ ~&~ [38.3, 62.0]      ~&~  $0.79_{-0.29}^{+0.18}$  ~&~ [0.35, 0.98]  ~&~  $116.2_{-13.4}^{+17.2} $  ~&~ [91.5, 142.5] \\ \hline
        GW200129\_065458 ~&~ 3   ~&~ $32.0_{-2.6}^{+1.7}        $ ~&~ [30.0, 45.2]      ~&~  $0.85_{-0.40}^{+0.12}$  ~&~ [0.26, 0.98]  ~&~  $74.6_{-3.8}^{+4.4}    $  ~&~ [67.0, 98.0] \\ \hline
        GW200208\_130117 ~&~ 3   ~&~ $38.8_{-4.7}^{+5.2}        $ ~&~ [32.7, 53.6]      ~&~  $0.73_{-0.29}^{+0.23}$  ~&~ [0.36, 0.98]  ~&~  $91.4_{-10.0}^{+11.4}  $  ~&~ [78.4, 127.1] \\ \hline
        GW200219\_094415 ~&~ 3   ~&~ $43.7_{-6.2}^{+6.3}        $ ~&~ [23.7, 51.8]      ~&~  $0.76_{-0.31}^{+0.20}$  ~&~ [0.23, 0.98]  ~&~  $102.5_{-11.9}^{+14.1} $  ~&~ [61.7, 131.3] \\ \hline
        GW200224\_222234 ~&~ 3   ~&~ $40.9_{-3.8}^{+3.5}        $ ~&~ [34.6, 57.0]      ~&~  $0.82_{-0.26}^{+0.15}$  ~&~ [0.28, 0.98]  ~&~  $94.8_{-7.1}^{+8.2}    $  ~&~ [82.8, 129.8] \\ \hline
        GW200225\_060421 ~&~ 3   ~&~ $17.6_{-1.9}^{+0.9}        $ ~&~ [15.3, 24.8]      ~&~  $0.72_{-0.27}^{+0.23}$  ~&~ [0.27, 0.98]  ~&~  $41.2_{-4.0}^{+2.9}    $  ~&~ [36.5, 57.9] \\ \hline
        GW200302\_015811 ~&~ 3   ~&~ $29.8_{-4.1}^{+7.3}        $ ~&~ [25.9, 46.4]      ~&~  $0.53_{-0.20}^{+0.35}$  ~&~ [0.30, 0.98]  ~&~  $73.7_{-7.9}^{+14.8}   $  ~&~ [63.1, 106.8] \\ \hline
        GW200311\_115853 ~&~ 3   ~&~ $32.6_{-2.7}^{+2.6}        $ ~&~ [24.6, 38.1]      ~&~  $0.81_{-0.26}^{+0.16}$  ~&~ [0.27, 0.98]  ~&~  $75.8_{-5.6}^{+6.1}    $  ~&~ [57.5, 93.7] \\ \hline
        GW200316\_215756 ~&~ 3   ~&~ $10.6_{-0.1}^{+0.1}        $ ~&~ [10.3, 11.2]      ~&~  $0.59_{-0.38}^{+0.34}$  ~&~ [0.01, 0.98]  ~&~  $25.5_{-1.0}^{+8.7}    $  ~&~ [21.7, 85.6] \\
        \hline
    \end{tabular}
 \caption{\ac{NN} model tests on \ac{GWTC} events. The Event\_id column lists the \ac{GWTC} event identifiers. The GWTC column denotes the version of the catalog from which the median $M_{\rm c}^{\rm PE}$, $q^{\rm PE}$, and $M_{\rm tot}^{\rm PE}$ parameters are obtained. The \ac{NN} intervals are computed with 96\% target accuracy.}
\label{tab:real_events}
\end{table*}
\end{center}

%% file: prior.tex
\section{NN model outputs in low-latency PE}\label{pe_analysis}

\texttt{Bilby} \cite{bilby}, is a Bayesian inference algorithm that employs the nested sampling technique with the Dynesty library implementation \cite{dynesty} to explore the full parameter space of masses and spins. Bilby's configuration in \ac{O4} takes into account uncertainties in detector calibration and marginalizes the posterior probability distribution over them. A \ac{ROQ} technique \cite{Canizares_roq, smith_roq, soichiro_roq} is implemented to reduce the computational cost of likelihood evaluations. Bilby determines the lower and upper bounds of chirp mass and mass ratio priors by estimating the pipeline uncertainty in recovering the template's chirp mass. The width of this priors affects the convergence of the algorithm and the speed of the \ac{PE} process. Since the \ac{NN} can provide accurate narrower intervals compared to Bilby's default choices, it could be a viable alternative for the first stage of the \ac{PE} process. 

In order to test this, we select 34 \ac{BBH}, \ac{NSBH}, and \ac{BNS} events each with the highest \ac{SNR} in two weeks of the O3 Replay MDC. We assume the boundary between \ac{NS} and \ac{BH} to be $3~M_\odot$. We run our models with a target accuracy of 98.8\% and 96\% for the mass ratio and the chirp mass, respectively. The higher target accuracy for $q$ is justified by the fact that in the low-mass region of the parameter space $M_{\rm c}$ is well estimated by the pipeline. Therefore, a larger interval for $q$ is required to maximize the range of component masses searched by \ac{PE}.

We produce full posteriors with \verb|bilby 2.2.1| and \verb|bilby-pipe 1.3.0| and uniform mass priors with intervals computed by the \ac{NN}. We also perform \ac{PE} on the same set with the default low-latency \texttt{bilby}  listed in Table \ref{tab:bilby_prior_range}. We employ the relative binning likelihood \cite{relative_binning} since the low-latency \ac{ROQ} method is incompatible with the dynamical chirp mass intervals provided by the \ac{NN}. 

Low-latency \texttt{Bilby} runs with pre-defined chirp mass intervals set by the \ac{ROQ} method. Because of this restriction, the \ac{PE} process cannot  initiate unless the \ac{NN} estimates lie in these pre-defined intervals.

\begin{table}[h]
\renewcommand{\arraystretch}{1.18}
\centering
    \begin{tabular}{|c|c|c|}
    \hline
    \multirow{2}{*}{\textbf{Conditions}} & \multicolumn{2}{c|}{\textbf{Prior bounds on $M_{\rm c}$}}  \\ \cline{2-3}
      & \multicolumn{1}{c|}{Lower} & \multicolumn{1}{c|}{Upper}    \\ \hline
      $M_{\rm c} <$  2  &  $M_{\rm c} - 0.01$ & $M_{\rm c}$ + 0.01 \\ \hline
      $ 2 < M_{\rm c} <$  4  &  $M_{\rm c} - 0.1$ & $M_{\rm c}$ + 0.1 \\ \hline
      $ 4 < M_{\rm c} <$  8  &  $M_{\rm c} \times 0.9$ & $M_{\rm c} \times 1.1$ \\ \hline
      $ 8 < M_{\rm c} <$  9.57  &  5.21 & 10.99 \\ \hline
      $ 9.57 < M_{\rm c} <$ 12  &  8.71 & 20.99 \\ \hline
      $ 12 < M_{\rm c} <$ 16  &  10.03 & 19.04 \\ \hline
      $ 16 < M_{\rm c} <$ 25  &  12.8 & 31.8 \\ \hline
      $ 25 < M_{\rm c} <$ 45  &  18.8 & 62.8 \\ \hline
      $ M_{\rm c} > 45$   &  29.9 & 199.9 \\ \hline
    \end{tabular}
\caption{\texttt{Bilby}'s default prior bounds on the chirp mass for low-latency \ac{PE}. All masses are in units of solar mass. The prior on the mass ratio is uniformly distributed within [0.125,1] for $M_{\rm c} < 2 M_\odot$ and [0.06, 1] otherwise.}
\label{tab:bilby_prior_range}
\end{table}

In the nested sampling algorithm, each likelihood iteration replaces the lowest likelihood sample with a higher likelihood sample. This process efficiently converges on high-probability regions of parameter space. Narrow priors around the injected value typically lead to a faster convergence of the algorithm. Although this may produce posterior distributions with limited support, the method is valuable in low-latency applications where greater speed is preferable.

\begin{figure}
    \centering
    \includegraphics[width=\linewidth]{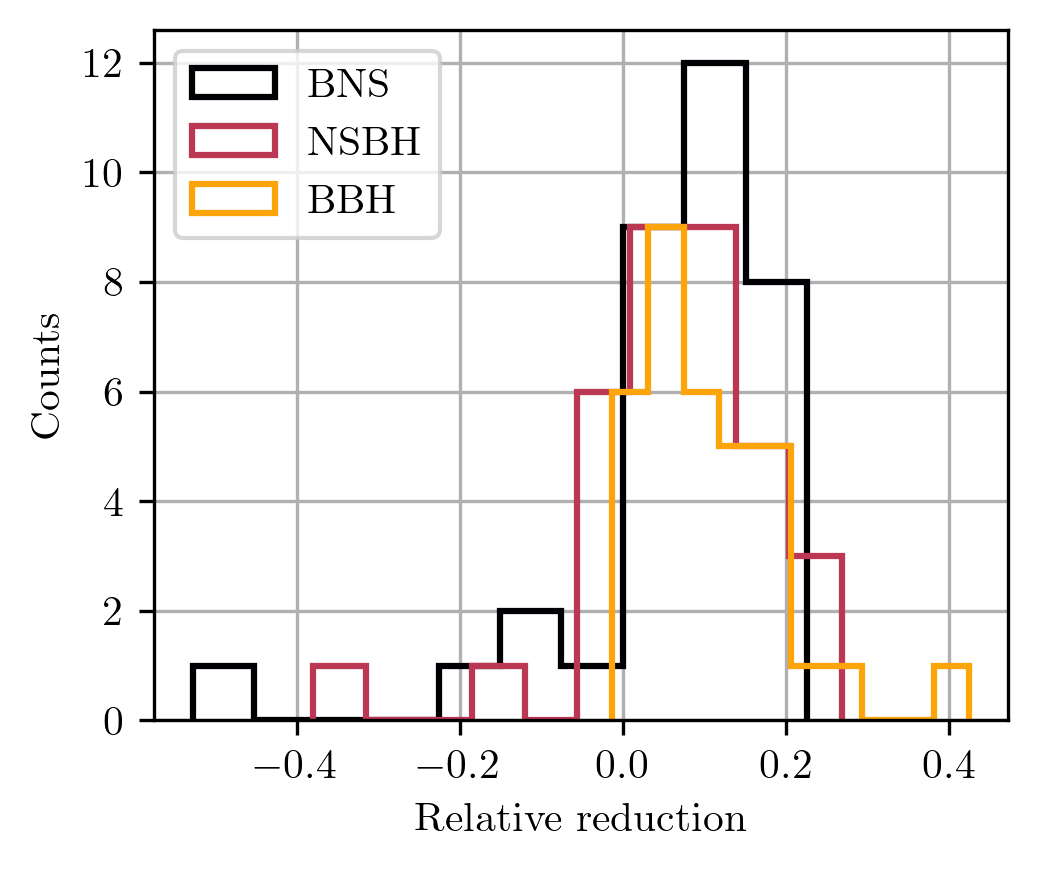}
    \caption{Relative reduction in the number of likelihood iterations between \ac{NN} and default \texttt{Bilby}. The median gain is 0.10, 0.07, and 0.09 for \ac{BNS}, \ac{NSBH}, and \ac{BBH} sets, respectively.}
    \label{fig:likelihoodevals}
\end{figure}

Figure \ref{fig:likelihoodevals} compares the number of likelihood evaluations for \texttt{Bilby}'s default configuration and with the \ac{NN} priors. The \ac{NN} priors show a $\sim 9$\% median reduction in the number of likelihood evaluations. Such a reduction is significant in the low-latency scenario, where \ac{PE} may still require a few hours time even for well localized \ac{BBH}s.

\begin{figure*}
    \centering
    \includegraphics[width=0.48\linewidth]{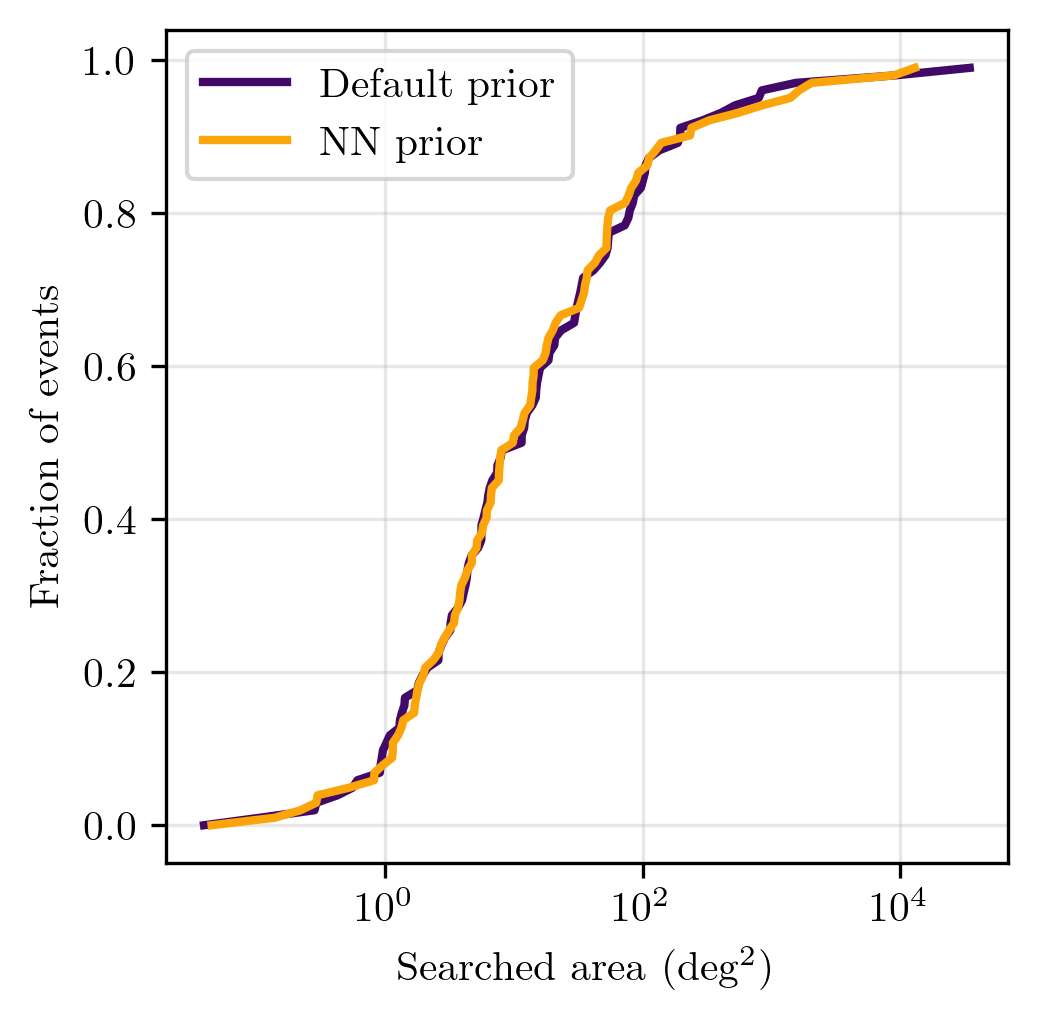}
    \includegraphics[width=0.48\linewidth]{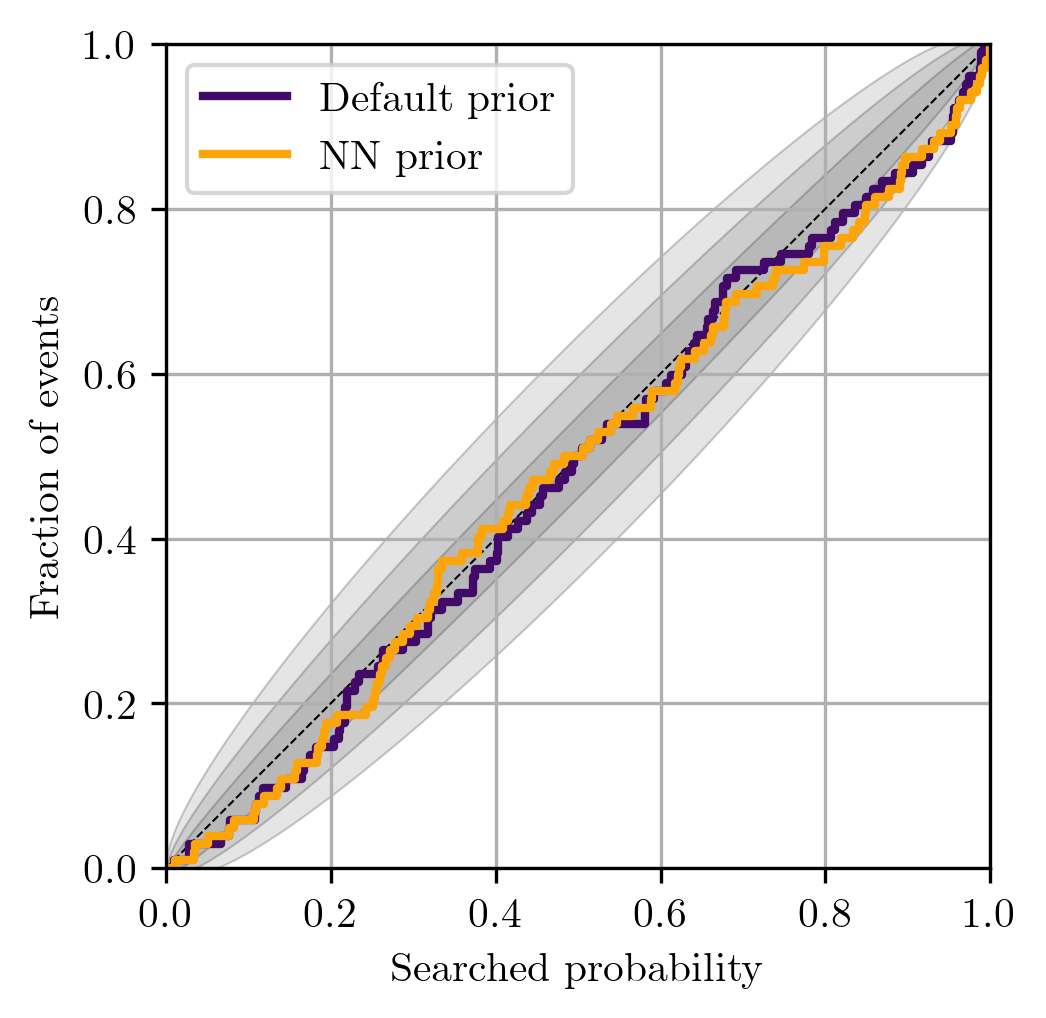}
    \caption{Skymap statistics for the selected MDC events. $Left:$ Cumulative searched area plots for the selected MDC events.
    $Right$ PP-plot showing the skymap statistics for the selected MDC events. The credible intervals shown
    in gray are based on the total number of events.}
    \label{fig:skyloc-searchedarea}
\end{figure*}

As a final check, we compare the event sky maps produced with \ac{NN} and default priors. The searched area event distribution and the corresponding p-p plot are shown in Fig.~\ref{fig:skyloc-searchedarea}. Both \ac{NN} and default prior are consistent.

%% file: conclusion.tex
\section{Conclusions}\label{sec:conclusions}

Low-latency \ac{PE} of \ac{GW} \ac{CBC} signals requires as an input accurate source parameter priors. We have presented a new \ac{NN} method based on \ac{QR} that yields confidence bounds for the chirp mass, mass ratio, and total mass source parameters of candidate events identified in online \ac{LVK} searches. The convergence of the \ac{PE} process relies on the width of the priors; the narrower the prior interval, the faster the \ac{PE} calculation converges. The \ac{NN} \ac{QR} provides accurate estimates for chirp mass, mass ratio, and total mass intervals that are comparable to the errors from the search pipelines in recovering these parameters. Therefore, our method provides dynamic intervals narrower than those currently in use that still enable accurate low-latency \ac{PE} while speeding up the process. It could be used to provide confidence bounds of event intrinsic parameters in real time, rather than point estimates that may suffer from search pipeline bias.

We train the \ac{NN} on data from the second \ac{LV} observing run and test it on \ac{O2} data, as well as data from the \ac{O3} \ac{MDC} dataset and real events from \ac{GWTC}. The method performs consistently across all datasets with an accuracy of over 90\%. 

The algorithm performance does not affect the quality of the event sky localizations; the searched area and searched probability derived from the \ac{NN} priors are consistent with the current \ac{LVK} implementation.

In our analysis we applied the \ac{NN} method to the Bilby \ac{PE} pipeline. The \ac{LVK} is currently employing a second pipeline for low-latency \ac{PE}. \texttt{RAPIDPE-RIFT} \cite{rapid_pe} is an algorithm that utilizes a non-Markovian sampler. The pipeline fixes the \ac{CBC} intrinsic parameters on a grid and marginalizes over the remaining parameters to speed up the process. The initial grid region is determined from the template parameters. Therefore, the \ac{NN} could be utilized to constrain the parameter ranges in the initial grid. Since the number of points in \texttt{RAPIDPE-RIFT}'s grid is fixed, a \ac{NN} implementation would not decrease the latency of the pipeline. However, narrower intervals combined with \texttt{RAPIDPE-RIFT}'s adaptive mesh refinement \cite{caitlin_adaptive_mesh} would increase the precision of the \ac{PE}. 

Our method could be further improved by training it on \ac{O4} datasets with better coverage of the parameter space. This approach would help the model increase its parameter space support and keep up to date with pipeline updates. Another refinement could be to extend the \ac{NN} input to multiple independent pipelines and parameters not restricted to the highest \ac{SNR} template.

%The NN has also proven its utility in setting up priors for low-latency \ac{PE} algorithms. Given its accuracy and narrow predictive intervals, it can be used to provide more information to astronomers in real time (e.g. $M_c$ and $M_{tot}$ intervals) for \ac{EM} follow-up purposes as well.

%The \ac{QR} algorithm allows for the definition of the output target accuracy before training. 

\section{Acknowledgments}

This work is based upon work supported by the LIGO Laboratory which is a major facility fully funded by the U.s.\ National Science Foundation. We are grateful for computational resources provided by the LIGO Laboratory
and supported by the U.S.\ National Science Foundation Awards PHY-0757058 and PHY-0823459. We express our gratitude to the many colleagues from the \ac{LVK} collaborations, without whom this work would not have been possible. In particular, we thank Daniel Wysocki, Carl-Johan Haster, and Soichiro Morisaki for their help and productive comments. G.P.\ thanks Giovanni Andrea Prodi for useful discussions and the supervision of his master thesis.

S.S.C.\ is partially supported by the U.S.\ National Science Foundation under awards PHY-2011334 and PHY-2308693. This research was initiated during G.P.'s stay at the Institute of Multi-messenger Astrophysics and Cosmology, Missouri University of Science and Technology, as part of the 2023 INFN-NSF/LIGO Summer Exchange Program. G.P.\ acknowledges support from the Italian INFN (Istituto Nazionale di Fisica Nucleare) with scholarships n.\ 25384/2023 and 25912/2023. M.C.\ is partially supported by the U.S. National Science Foundation under awards PHY-2011334, PHY-2219212 and PHY-2308693. 

This manuscript has been assigned LIGO Document Control Center number P2500183.

%% file: appendix.tex
%\appendix

%Similar results are achieved on the O3 MDC replay dataset, not only for the $M_c$ models but also for the $q$ and $M_{tot}$ models.

\begin{figure*}[!htp]
    \centering 
    \includegraphics[width=0.95\linewidth]{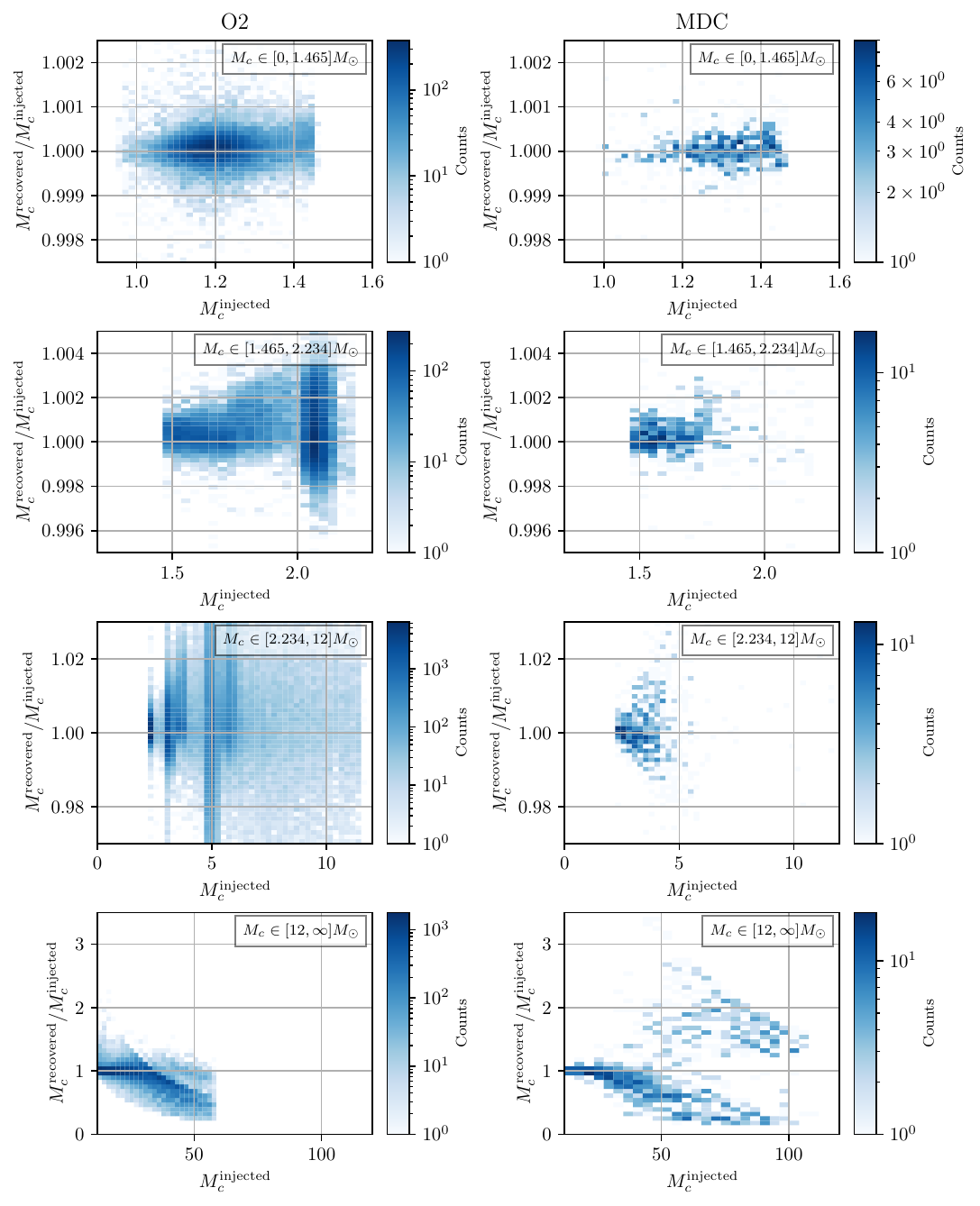}
    \caption{
        Comparison between the recovered and injected chirp masses in the O2 and O3 replay MDC dataset. The y-axis represents the ratio of pipeline recovered $M_c$ to injected $M_c$. The pipeline behavior in scenarios with  $M_{\rm c}^{\rm inj}\gtrsim 30 \si{\msun}$ (last row) has remained similar between O2 and O3 replay MDC dataset. The pipelines tend to underestimate the true value in this region. In the O2 dataset, there were no injections higher than $\SI{60}{\msun}$ but are present in the MDC data.  In these cases the pipeline can  overestimated the values by a factor larger than 2.
    }
    \label{fig:MDCvsO2_Mc}
\end{figure*}
\begin{figure*}[!htp]
    \centering
    \includegraphics[width=0.95\linewidth]{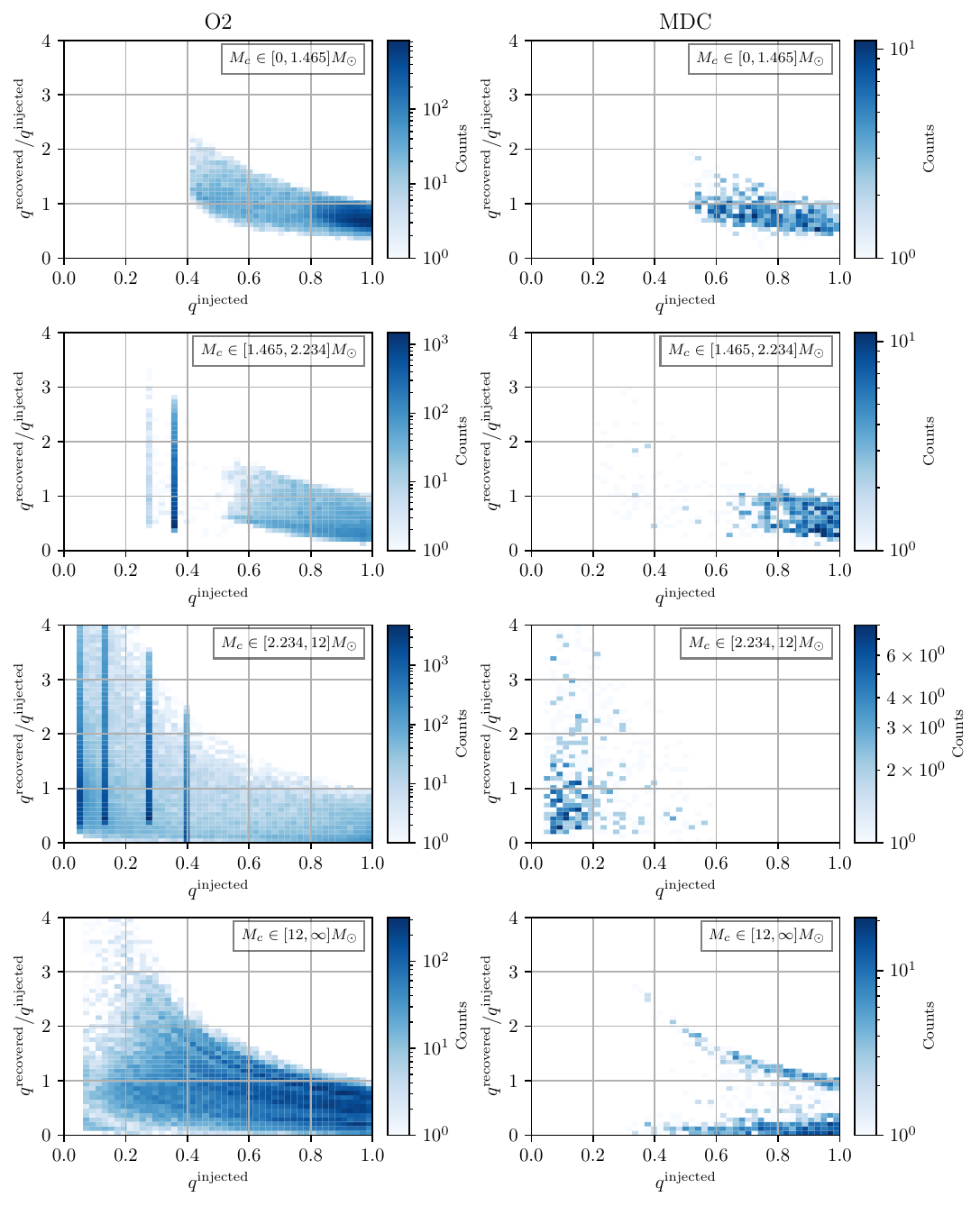}
    \caption{Comparison between the recovered and the injected mass ratio in the O2 and MDC datasets.  The quality of the recovery is much worse than for the chirp mass, with relative errors of order unity in all bins. This happens because $q$, unlike the chirp mass, does not enter the waveform at first order.
    }
    \label{fig:MDCvsO2_q}
\end{figure*}
\begin{figure*}[!htp]
    \centering
    \includegraphics[width=0.95\linewidth]{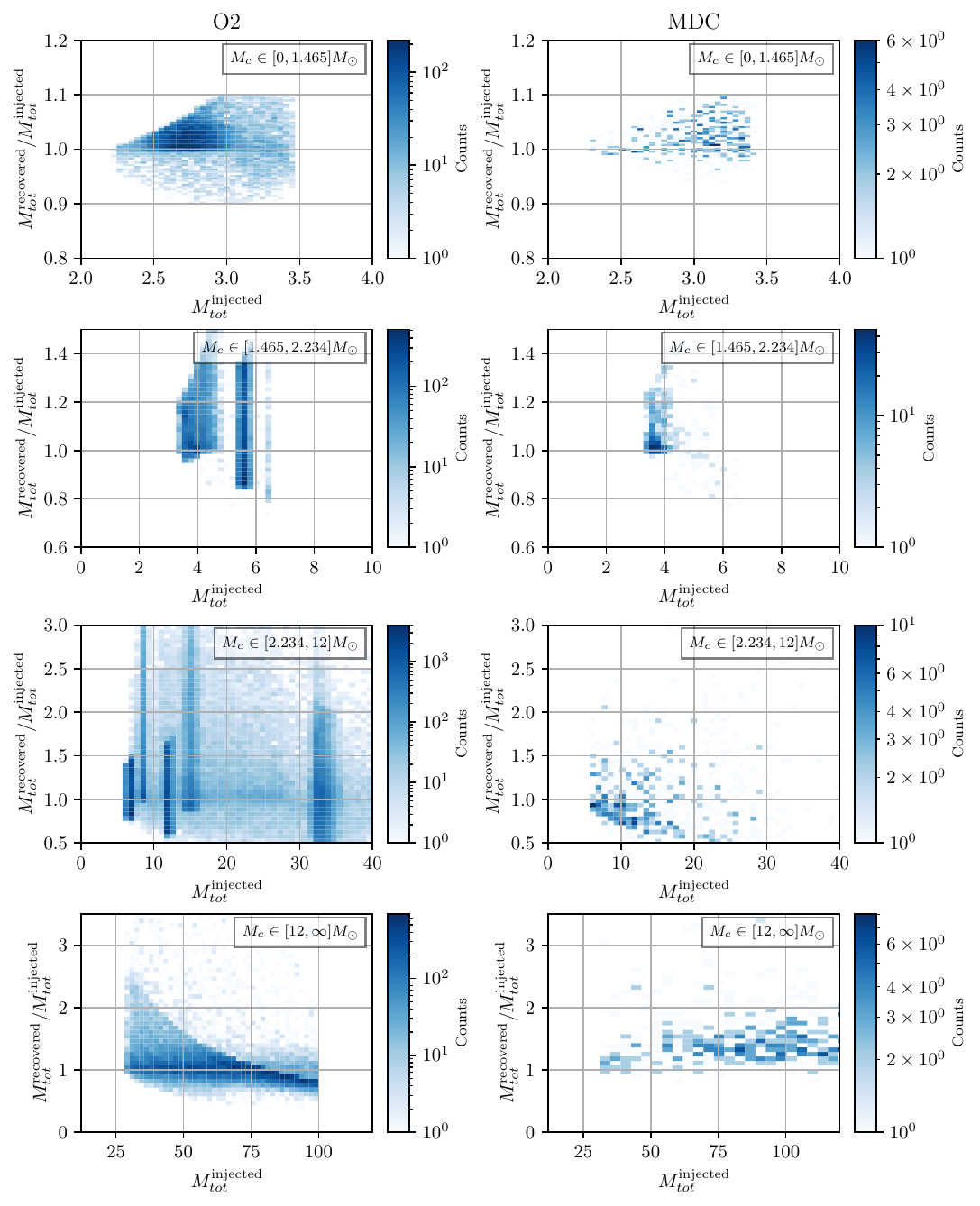}
    \caption{Comparison between the recovered and injected total mass in the O2 and MDC datasets.  we see that O2 includes a much larger range of injected total mass, especially on bin C.
    As in the case of the mass ratio, the recovery quality is worse than for the chirp mass.}
    \label{fig:MDCvsO2_Mtot}
\end{figure*}

%\section{Mass ratio accuracy}
%As discussed in section \ref{sec:results_O2},  in the $q$ model the \ac{NN} output is passed through a sigmoid function which limits the \ac{NN} output $\in (0,1)$. This causes some unwanted artifact in accuracy plots since equal mass ratio events ($q$=1) will always be classified as inaccurate. However, if we assume a $q$ threshold  at 0.98 such that if $q > 0.98$ set $q=1$, then we can see  the changes in fig \ref{fig:q_threshold}. Bin A accuracy at higher target accuracies becomes closer to one with the threshold compared to accuracy without threshold. The curve is above the diagonal because the threshold induces some bias which causes the measured accuracy to be larger than the target accuracy. Bin C however shows little to no change with a maximum accuracy below 95\%. Bin C has the highest amount of among all other bins but contains multiple delta function-like peaks in the injected distribution which is not limited to only $q$. These dense injections are all less than $q <0.5$. 
%\begin{figure}[t]
%    \centering
%    \includegraphics[width=0.9\linewidth]{new_plots/thresholded_accuracy_bilby_O2 injection.pdf}
%    \caption{The accuracy of the $q$ model across different bins with and without $q$ threshold on the testing set.}
%    \label{fig:q_threshold}
%\end{figure}